\documentclass[12pt]{article}
\usepackage{amssymb,amsmath,bm,bbm}
\usepackage{epsf}
\usepackage{epsfig}
\usepackage{afterpage}
\usepackage{longtable}
\usepackage{cite}
\usepackage{latexsym, mathrsfs}
\usepackage{axodraw}
\usepackage{graphics}
\usepackage{color}

\DeclareMathOperator{\diag}{diag}

\def\slash#1{\setbox0=\hbox{$#1$}\dimen0=\wd0
      \setbox1=\hbox{/} \dimen1=\wd1 \ifdim\dimen0>\dimen1
      \rlap{\hbox to \dimen0{\hfil/\hfil}} #1                        \else
      \rlap{\hbox to \dimen1{\hfil$#1$\hfil}}
      /   \fi}

\newcommand{\lsim}{
\mathrel{\hbox{\rlap{\hbox{\lower4pt\hbox{$\sim$}}}\hbox{$<$}}}}

\newcommand{\gsim}{
\mathrel{\hbox{\rlap{\hbox{\lower4pt\hbox{$\sim$}}}\hbox{$>$}}}}

\newcommand{\tev}{\, {\rm TeV}}
\newcommand{\gev}{\, {\rm GeV}}

\def\beq{\begin{equation}}
\def\eeq{\end{equation}}

\newcommand{\be}{\begin{equation}}
\newcommand{\ee}{\end{equation}}
\newcommand{\bea}{\begin{eqnarray}}
\newcommand{\eea}{\end{eqnarray}}

\newcommand{\bi}{\begin{itemize}}
\newcommand{\ei}{\end{itemize}}

\newcommand{\newsection}[1]{\section{#1}\setcounter{equation}{0}}

\definecolor{gray}{rgb}{.38,.38,.38}

\begin{document}
\begin{titlepage}
\vspace*{-0.5truecm}

\begin{flushright}
{TUM-HEP-745/09}
\end{flushright}

\vspace{2cm}
\begin{center}
\boldmath
{\Large\textbf{A Comparative Study of Contributions to $\epsilon_K$\\\vspace{.2cm} in the RS Model}}
\unboldmath
\end{center}

\vspace{0.3truecm}

\begin{center}
{\bf Bjoern Duling$^a$}
\vspace{0.4truecm}

{\footnotesize
 $^a${\sl Physik Department, Technische Universit\"at M\"unchen,
D-85748 Garching, Germany}\vspace{0.2cm}
}

\end{center}

\vspace{2cm}
\begin{abstract}
\noindent 
We contrast the impact of Higgs mediated flavor changing neutral currents on $\epsilon_K$ in the framework of a warped extra dimension that was recently calculated by Azatov et al.~with the older results for Kaluza-Klein gluon induced corrections to that observable. We find that 
the most stringent constraint on the KK scale for a Higgs field localized on the infrared brane {for reasonable additional assumptions} comes from KK gluon exchange. In the case of a bulk Higgs field we show that for certain scenarios the Higgs contribution can in fact exceed the KK gluon contribution. In the course of this analysis we also describe in detail the different renormalization procedures that have to be employed in the KK gluon and Higgs cases to relate the new physics at high energies to low energy observables.
\end{abstract}

%
%
%
\end{titlepage}

\setcounter{page}{1}
\pagenumbering{arabic}

\newsection{Introduction}
Recently, Higgs flavor changing neutral currents (FCNCs) in Randall-Sundrum \cite{Randall:1999ee} (RS) models have received a lot of attention. A model independent analysis was performed in \cite{Agashe:2009di} (see also \cite{Low:2009di}) and it was shown that a light composite Higgs that couples strongly to new heavy states can lead to significant bounds on the compositeness scale from $K^0-\bar K^0$ oscillations. In \cite{Blanke:2008zb,Buras:2009ka}, where only the first fermionic Kaluza-Klein (KK) excitations in the custodially protected RS model (RSc) were considered it was found that tree-level Higgs FCNCs are negligible. However, in contrast to this in \cite{Azatov:2009na} it was pointed out that the summation over the whole KK tower of fermionic excitations yields a finite contribution to Higgs FCNCs. In the same paper also the resulting Higgs contribution to $\epsilon_K$ was calculated and a Higgs-mass-dependent bound on the KK mass scale was deduced. 
While the strongest bound on the KK mass scale in RS models with a brane Higgs is due to the tree-level exchange of KK gluons \cite{Csaki:2008zd,Blanke:2008zb}, the bound deduced in \cite{Azatov:2009na} is well of the same order of magnitude, at least for a light Higgs. 
In view of these new results it is mandatory to compare the bounds on the KK scale that are required to keep under control the effects of the tree level exchanges of KK gluons
on the one hand and of the Higgs boson on the other hand. In particular this is necessary since the bounds of \cite{Csaki:2008zd,Blanke:2008zb} and \cite{Azatov:2009na} are not comparable to each other without further ado as they depend on the authors' prejudices towards naturalness or the experimental and theoretical uncertainties in $\epsilon_K$. Even more so these assumptions are implicit in \cite{Azatov:2009na}.

The rest of this paper is organized as follows. In Section \ref{sec:Higgs-FCNCs} we briefly review the results for the flavor off-diagonal Higgs couplings derived in \cite{Azatov:2009na} and comment on their renormalization group (RG) evolution. Section \ref{sec:epsK} is devoted to the calculation of $\epsilon_K$ in terms of the flavor off-diagonal Higgs couplings in the brane-Higgs scenario. In Section \ref{sec:Numerical_Results} we compare our results for the Higgs and KK gluon contributions in the brane Higgs scenario and show how this result can be generalized to the case of a bulk Higgs.
Our conclusions finally are presented in Section \ref{sec:Conclusions}.

\begin{figure}[htbp]
\begin{center}
\includegraphics[height=2.5cm]{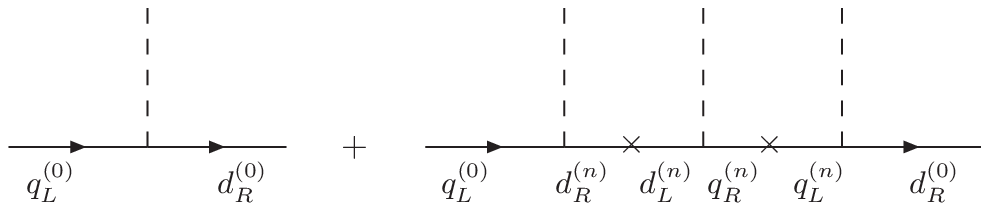}
\end{center}
\caption{RS contributions to quark masses and Yukawa couplings.\label{fig:MIA}}
\end{figure}

\newsection{Flavor-changing Higgs couplings}\label{sec:Higgs-FCNCs}
In this section we briefly recapitulate the main results of \cite{Azatov:2009na} and comment on the RG evolution of the flavor off-diagonal Higgs couplings. A detailed description of the model setup as well as the notation used in the present paper can be found in \cite{Albrecht:2009xr}. To set some additional notation we explicitly write out the relevant Lagrangian
\begin{equation}
\mathcal L_\text{Yuk}=\sum\limits_{n_1=0}^\infty\bar q_L^{i(n_1)}\left[\hat Y_1\right]_{ij}\sum\limits_{n_2=0}^\infty d_R^{j(n_2)}H
+\sum\limits_{m_1=1}^\infty\bar d_L^{i(m_1)}\left[\hat Y_2\right]_{ij}\sum\limits_{m_2=1}^\infty q_R^{j(m_2)}H+h.c.\,,
\label{eq:L_Yuk}
\end{equation}
where $\hat Y_1$ and $\hat Y_2$ are fundamental 5D Yukawa matrices. Flavor off-diagonal Higgs couplings in the mass eigenbasis can arise whenever the RS contributions to quark masses and to the Yukawa couplings are not aligned. In the mass insertion approximation the RS contributions up to $\mathcal O(v^2/M_\text{KK}^2)$ are represented by the two diagrams in Fig.~\ref{fig:MIA} with the couplings given in (\ref{eq:L_Yuk}). The first diagram in Fig.~\ref{fig:MIA} contributes equally to the quarks' masses after EWSB and their Yukawa couplings. The second diagram however affects masses and Yukawa couplings in a different manner since its contribution to the Yukawa couplings comes with a combinatorial factor of three that is due to the three different choices of which two external Higgs lines are set to their VEVs. This shift between quark masses and Yukawa couplings results in flavor off-diagonal Higgs couplings once we go to the mass eigenstate basis.
At a first glance the overall contribution from the second diagram in Fig.~\ref{fig:MIA} seems to be negligible since both the $q_R^{(n)}$ and $d_L^{(n)}$ modes obey Dirichlet boundary conditions on the IR brane. In \cite{Azatov:2009na} the point has been made that the profiles of $q_R^{(n)}$ and $d_L^{(n)}$ do not exactly vanish on the IR brane but display a small discontinuity that is proportional to the Higgs VEV. After regularization of this discontinuity and summing over the infinite tower of KK modes it is found that a non-vanishing misalignment between quark masses and Yukawa couplings is generated by this diagram. At this point it is appropriate to call the reader's attention to the fact that the Yukawa matrix $\hat Y_2$ that couples the scalar currents $d_L^{i(n)}q_R^{j(n)}$ to the Higgs field is not required for the generation of quark masses and hence could be set to zero which would eliminate the second diagram's contribution to flavor off-diagonal Higgs couplings. However, since this choice for $\hat Y_2$ without profound physical reason contradicts naturalness, in the following we will set $\hat Y_2$ to be equal\footnote{Note that the choice $\hat Y_2=\hat Y_1$ is mandatory in the bulk Higgs scenario.} to $\hat Y_1$.

An additional source of misalignment between quark masses and Yukawa couplings is the modification of the kinetic terms by the mixing of SM quarks and KK quarks after EWSB as was first pointed out in \cite{Casagrande:2008hr} (see also \cite{Bauer:2009cf}). These flavor-dependent corrections to the kinetic terms make redefinitions of the quark fields necessary which in turn give rise to an additional shift between quark masses and Yukawa couplings. For the first two generations of quarks this contribution is found to be negligible, for the third generation however this effect can be of the same size as the one outlined above.\\

After this rather qualitative description we now summarize the main results of \cite{Azatov:2009na} for the case of a brane-localized Higgs field. The total misalignment between quark masses and Yukawa couplings comprises two contributions,
$\hat\Delta^d=\hat\Delta_1^d+\hat\Delta_2^d$, where $\hat\Delta^d_1$ is the contribution represented by the diagrams in Fig.~\ref{fig:MIA} and $\hat\Delta^d_2$ is due to rescaling of the quark fields as to canonize their kinetic terms. Explicitly, the authors of \cite{Azatov:2009na} find
\begin{equation}
\hat\Delta_1^d=\frac{2}{3}\hat F_Q\hat Y_1^{5D}\left(\hat Y_2^{5D}\right)^\dagger \hat Y_1^{5D}\hat F_d\frac{v^3}{f_{RS}^2}\,,\label{eq:Delta-d-1}
\end{equation}
and
\begin{equation}
\hat \Delta_2^d=\hat m^d\left(\hat m^{d\dagger}\hat K(c_Q)+\hat K(-c_d)\hat m^{d\dagger}\right)\hat m^d\frac{1}{f_\text{RS}^2}\,,
\end{equation}
where here and in the following a hat indicates a $3\times3$ matrix in flavor space. $f_\text{RS}=ke^{-kL}\approx M_\text{KK}/2.45$ is the warped-down curvature of the extra dimension that sets the scale of mass of the lightest KK states. The matrices $\hat K(c)=\diag K(c^i)$ and $\hat F_{Q,d}=\diag f(c^i_{Q,d})$ are functions that depend on the quark localization and are defined via
\begin{eqnarray}
f(c)&\equiv&\sqrt{\frac{1-2c}{1-e^{-(1-2c)kL}}}\,,\\
K(c)&\equiv&\frac{1}{1-2c}\frac{1}{e^{(1-2c)kL}-1}\left(-1+\frac{e^{(1-2c)kL}-e^{-2kL}}{3-2c}+\frac{e^{(2c-1)kL}-e^{-2kL}}{1+2c}\right)\,. 
\end{eqnarray}

The flavor off-diagonal components of the Yukawa couplings in the mass eigenbasis are obtained from $\hat\Delta^d$ via the bi-unitary transformation
\begin{equation}
\hat Y_\text{off-diag.}=\mathcal D_L^\dagger\hat\Delta^d\mathcal D_R\,,
\end{equation}
where $\mathcal D_L$ and $\mathcal D_R$ are the unitary rotations that diagonalize the down-type quark mass matrix,
\begin{equation}
\hat m_d^\text{diag}=\mathcal D_L^\dagger\hat m_d\mathcal D_R\,.
\end{equation}\\

The misalignment between quark masses and the resulting flavor off-diagonal Yukawa couplings given above are generated at or beyond the KK mass scale.
In the following, we will take this scale to be $\mu_s=3\tev$. From this high energy scale the Yukawa couplings have to be evolved 
down to the scale $\mu_H$ of the Higgs mass where new effective interactions are generated by tree level exchanges of the Higgs boson.
The RG evolution of these couplings is in fact identical to that of quark masses and is well known to next-to-leading order (NLO). From this scale the Wilson coefficients of the new operators then have to be evolved down to the physically relevant scale $\mu_L=2\gev$ according to their individual anomalous dimensions. These RG effects have been neglected in recent publications discussing Higgs FCNCs, e.g.~\cite{Giudice:2008uua}. For a complete numerical analysis however these effects have to be included. We will discuss this issue in the following section.

\boldmath\newsection{Higgs contributions to $\epsilon_K$}\unboldmath\label{sec:epsK}
The Lagrangian relevant for Higgs contributions to $\Delta S=2$ transitions in the quark mass eigenbasis is given by
\begin{equation}
\mathcal L_\text{NC}^\text{Higgs}=-\hat Y_{21}\bar s_L d_R H-\hat Y_{12}^\ast\bar s_R d_L H\,,
\end{equation}
where $\hat Y$ is the down-type $3\times3$ Yukawa matrix for quarks in the mass eigenbasis at energy scale $\mu=M_H$. If we define $\Delta_R^{H}\equiv\hat Y_{21}$, $\Delta_L^{H}\equiv\hat Y_{12}^\ast$, the effective Hamiltonian for $\Delta S=2$ transitions that are induced by tree-level Higgs exchanges is found to be
\begin{eqnarray}
\left[\mathcal H_\text{eff}^{\Delta S=2}\right]_\text{Higgs}&=&\frac{1}{2M_H^2}\left[\left(\Delta_L^H\right)^2(\bar s P_L d)(\bar s P_L d)+\left(\Delta_R^H\right)^2(\bar s P_R d)(\bar s P_R d)\right.\nonumber\\&+&2\Delta_L^H \Delta_R^H(\bar s P_L d)(\bar s P_R d)\Big]\,,\label{eq:H_eff1}
\end{eqnarray}
where $P_{L,R}=\frac{1}{2}(1\mp\gamma_5)$. Here and in the following summation over color indices within the brackets of the effective operators is understood. In the operator basis of \cite{Buras:2001ra} this is equivalent to the effective Hamiltonian
\begin{equation}
\left[\mathcal H_{eff}^{\Delta S=2}\right]_{Higgs}=\frac{1}{2M_H^2}\left[\left(\Delta_L^H\right)^2Q_1^{SLL}+\left(\Delta_R^H\right)^2Q_1^{SRR}+2\Delta_L^H \Delta_R^HQ_2^{LR}\right]\,,\label{eq:H_eff2}
\end{equation}
and the Wilson coefficients at the energy scale $\mu=M_H$ are accordingly given by
\begin{eqnarray}
C_1^{SLL}(\mu_H)&=&\frac{1}{2M_H^2}\left(\Delta_L^H\right)^2\,,\label{eq:Wilsons1}\\
C_1^{SRR}(\mu_H)&=&\frac{1}{2M_H^2}\left(\Delta_R^H\right)^2\,,\label{eq:Wilsons2}\\
C_2^{LR}(\mu_H)&=&\frac{1}{M_H^2}\Delta_L^H\Delta_R^H\,.\label{eq:Wilsons3}
\end{eqnarray}

The effective interactions in (\ref{eq:H_eff2}) are generated at an energy scale $\mu=M_H$. From this scale the Wilson coefficients $C_i^X$ in (\ref{eq:Wilsons1})-(\ref{eq:Wilsons3}) have to be evolved down to the low energy scale $\mu=\mu_L=2\gev$ at which the hadronic matrix elements $\left\langle\bar K^0|Q_i^X|K^0\right\rangle$ of the operators in (\ref{eq:H_eff2}) are evaluated using lattice methods. This RG evolution can be performed separately from the additive SM contribution and the contributions from KK gauge bosons that have been discussed in \cite{Blanke:2008zb}. Under renormalization $Q_2^{LR}$ and $Q_1^{SLL,SRR}$ mix with $Q_1^{LR}$ and $Q_2^{SLL,SRR}$, respectively, so that the full operator basis relevant for the present analysis is given by
\begin{align}
 &Q_1^{LR}=(\bar s\gamma_\mu P_L d)(\bar s\gamma^\mu P_R d)\,,\\
 &Q_2^{LR}=(\bar s P_L d)(\bar s P_R d)\,,\\
 &Q_1^{SLL}=(\bar s P_L d)(\bar s P_L d)\,,\\
 &Q_2^{SLL}=(\bar s \sigma_{\mu\nu}P_L d)(\bar s \sigma^{\mu\nu}P_L d)\,,\\
 &Q_1^{SRR}=(\bar s P_R d)(\bar sP_R d)\,,\\
 &Q_2^{SRR}=(\bar s \sigma_{\mu\nu}P_R d)(\bar s \sigma^{\mu\nu}P_R d)\,,
\end{align}
and the RG evolution operators each are given by $2\times2$ matrices. It should be noted that due to the insensitivity of QCD to the sign of $\gamma_5$ the Wilson coefficients $C_1^{SLL}$ and $C_1^{SRR}$ evolve identically under the RG while their initial conditions (\ref{eq:Wilsons1}) and (\ref{eq:Wilsons2}) at scale $\mu_H$ are in general different from each other. 
Resorting to the expressions for the evolution operators given in \cite{Buras:2001ra} we find that all the Wilson coefficients (\ref{eq:Wilsons1})-(\ref{eq:Wilsons3}) are enhanced in the RG evolution from $\mu=M_H$ down to $\mu=\mu_L$ with the strongest enhancement occurring in the case of the $Q_2^{LR}$ operator.

From the effective Hamiltonian
at the low energy scale $\mu_L$,
\begin{eqnarray}
\left[\mathcal H_\text{eff}^{\Delta S=2}\right]_\text{Higgs}&=&C_1^{LR}(\mu_L)Q_1^{LR}+C_2^{LR}(\mu_L)Q_2^{LR}+C_1^{SLL}(\mu_L)Q_1^{SLL}+C_1^{SRR}(\mu_L)Q_1^{SRR}\nonumber\\&+&C_2^{SLL}(\mu_L)Q_2^{SLL}+C_2^{SRR}(\mu_L)Q_2^{SRR}\,,
\end{eqnarray}
the Higgs contribution to the off-diagonal element $M_{12}^K$ that is responsible for $K^0-\bar K^0$ oscillations is obtained by taking
\begin{equation}
 m_K\left(M_{12}^K\right)_\text{Higgs}^\ast=\langle\bar K^0|\left[\mathcal H_\text{eff}^{\Delta S=2}\right]_\text{Higgs}|K^0\rangle\,,
\end{equation}
with $m_K$ the neutral K-meson mass. The matrix elements $\left\langle\bar K^0|Q_i(\mu)|K^0\right\rangle\equiv\left\langle Q_i(\mu)\right\rangle$ can then be parameterized by
\begin{eqnarray}
\left\langle Q_1^{LR}(\mu)\right\rangle&=&-\frac{1}{6}R(\mu)m_K^2 F_K^2 B_1^{LR}(\mu)\,,\\
\left\langle Q_2^{LR}(\mu)\right\rangle&=&\frac{1}{4}R(\mu)m_K^2 F_K^2 B_2^{LR}(\mu)\,,\\
\left\langle Q_1^{SLL}(\mu)\right\rangle&=&-\frac{5}{24}R(\mu)m_K^2 F_K^2 B_1^{SLL}(\mu)\,,\\
\left\langle Q_2^{SLL}(\mu)\right\rangle&=&-\frac{1}{2}R(\mu)m_K^2 F_K^2 B_2^{SLL}(\mu)\,,
\end{eqnarray}
where
\begin{equation}
R(\mu)=\left(\frac{m_K}{m_s(\mu)+m_d(\mu)}\right)^2
\end{equation}
and $F_K$ is the K-meson decay constant. Since QCD is blind to the sign of $\gamma_5$, the matrix elements of $Q_{1,2}^{SRR}$ are identical to those of $Q_{1,2}^{SLL}$. The hadronic parameters $B_i^X$ are known from lattice calculations and are related to the parameters $B_2$, $B_3$, $B_4$ and $B_5$ calculated in \cite{Babich:2006bh} via
\begin{equation}
B_1^{LR}\equiv B_5\,,\quad B_2^{LR}\equiv B_4\,,\quad B_1^{SLL}\equiv B_2\,,\quad B_2^{SLL}\equiv\frac{5}{3}B_2-\frac{2}{3}B_3\,,
\end{equation}
and their numerical values at the lattice scale $\mu_L=2.0\gev$ are given by \cite{Babich:2006bh} 
\begin{equation}
B_2=0.679\,,\quad B_3=1.055\,,\quad B_4=0.810\,,\quad B_5=0.562\,.
\end{equation}

Finally, the CP-violating parameter in the $K^0-\bar K^0$ system is given by
\begin{equation}
\epsilon_K=\frac{\kappa_\epsilon e^{i\varphi_\epsilon}}{\sqrt{2}\left(\Delta M_K\right)_\text{exp}}\left[\textrm{Im}\left(M_{12}^K\right)_\text{SM}+\textrm{Im}\left(M_{12}^K\right)_\text{Higgs}\right]\,,
\end{equation}
where $\varphi_\epsilon=(43.51\pm0.05)^\circ$ and $\kappa_\epsilon=0.92\pm0.02$ \cite{Buras:2008nn} account for $\varphi_\epsilon\neq\pi/4$ and include an additional effect from the imaginary part of the 0-isospin amplitude in $K\to\pi\pi$.
The SM contribution $\left(M_{12}^K\right)_\text{SM}$ in the conventions used in the present paper can be found in \cite{Blanke:2008zb}.

\newsection{Numerical Results}\label{sec:Numerical_Results}
In our analysis we will proceed in the same manner as done in \cite{Blanke:2008zb}. In that paper the 28 parameters determining the fundamental Yukawa matrices $\hat Y_u^5$ and $\hat Y_d^5\equiv \hat Y_1$ are randomly chosen in their respective ranges, $[0,\pi/2]$, $[0,2\pi]$ and $[1/3,3]$ for angles, phases and absolute sizes of Yukawa couplings, respectively. The last range accounts for the fact that fundamental Yukawa couplings larger than about three\footnote{This bound corresponds to $Y_\text{KK}^\text{max}=6$ in \cite{Agashe:2008uz}.} result in a breakdown of perturbativity at energies below the mass of the second KK excitation (see eg.~\cite{Csaki:2008zd,Agashe:2008uz} for more details) and that too small values tend to require the right-handed top-quark to be extremely localized towards the infrared (IR) brane. The nine quark bulk mass parameters $c_{Q,u,d}^i$ are then determined such that the quark masses and CKM mixing angles are reproduced, which can be much facilitated by using the Froggatt-Nielsen formalism \cite{Froggatt:1978nt}. Finally, we keep only those parameter sets that in addition to the quark masses and CKM mixing angles also reproduce the proper value of the Jarlskog determinant, all within their respective $2\sigma$ ranges.

Since for given fundamental Yukawa matrices the choice of bulk mass parameters $c_{Q,u,d}^i$ is not unique due to the transformation
\begin{equation}
f(c^i_{Q})\to\zeta f(c^i_{Q})\,,\quad f(c^i_{u,d})\to\frac{1}{\zeta} f(c^i_{u,d})\,,\quad i=1,2,3\,,\quad \zeta\in\mathbbm{R}^+\,,\label{eq:shift-symmetry}
\end{equation}
as was pointed out in \cite{Casagrande:2008hr}, we additionally specify that in our scan $0.4\leq c_Q^3\leq0.45$ is taken.
However, it should be pointed out that the numerical consequences of modifying a given set of parameters according to (\ref{eq:shift-symmetry}) for $K^0-\bar K^0$ mixing are small for $\mathcal O(1)$ values of $\zeta$.\\

In the following we will scan the parameter space of the RSc model in the way described above. To make the statements of this paper fully traceable, we mention the three additional assumptions:
\begin{itemize}
 \item the fundamental Yukawa matrix $\hat Y_2$ is equal to $\hat Y_1$,
 \item the 5D QCD coupling constant at the KK scale is taken to by $g^\ast=g_s^{5D}\sqrt{k}=6$ and $g^\ast=3$,
 \item the Higgs field is localized at the IR brane ($\beta=\infty$).
\end{itemize}
In this, $\hat Y_2=\hat Y_1$ is considered to be a natural choice as this is mandatory in the case of a bulk Higgs. The value $g^\ast=6$ is obtained from tree level matching of the 5D to the 4D QCD coupling constants. In the case of loop level matching the minimal value $g^\ast=3$ is obtained, while gluonic brane kinetic terms could enhance $g^\ast$ beyond the tree level value \cite{Agashe:2008uz}. On the other hand, $g^\ast$ is bounded from above by perturbativity considerations. The requirement that the theory is perturbative beyond the second KK excitation results in a (conservative) upper bound of $g^\ast=3$, and $g^\ast=6$ is already disfavored by this estimate \cite{Agashe:2008uz}. To allow for a better comparison with the previous analyses \cite{Csaki:2008zd,Blanke:2008zb} we still include the higher value $g^\ast=6$ into our analysis.\\

Changes in $g^\ast$ will change the relative and absolute sizes of the KK gluon and Higgs contributions to $\epsilon_K$. Also, the hierarchy between gauge boson and Higgs contributions is affected by the localization of the Higgs field along the fifth dimension and the average absolute size $\langle\hat Y_1\rangle$ of the 5D Yukawa couplings.

More precisely, in the bulk Higgs scenario the overlap integral of the Higgs with the fermionic zero mode profiles also receives contributions from the fermions' bulk profiles. This allows to localize the fermion profiles closer to the UV brane which in turn decreases the overlap of the fermion modes with KK gauge bosons and thus diminishes the impact of KK gauge bosons on $\epsilon_K$ \cite{Agashe:2008uz,Gedalia:2009ws}.  On the other hand, as was argued in \cite{Azatov:2009na} the size of the flavor changing Higgs couplings is only mildly affected by the Higgs localization. In consequence, detaching the Higgs field from the IR brane increases the relative importance of the Higgs contribution.

Increasing the average absolute size $\langle\hat Y_1\rangle$ allows to move the fermion profiles closer to the UV brane, with the same effect as described above, but beyond that at the same time the Higgs contribution to $\epsilon_K$ is increased. Thus larger 5D Yukawa couplings also increase the relative importance of the Higgs contribution. We will investigate the impact of deviations of $g^\ast$, $\langle\hat Y_1\rangle$ and the Higgs localization parameter $\beta$ from their values chosen above in detail at the end of this section.\\

\begin{figure}[htbp]
\begin{center}
\includegraphics[height=.5\textwidth]{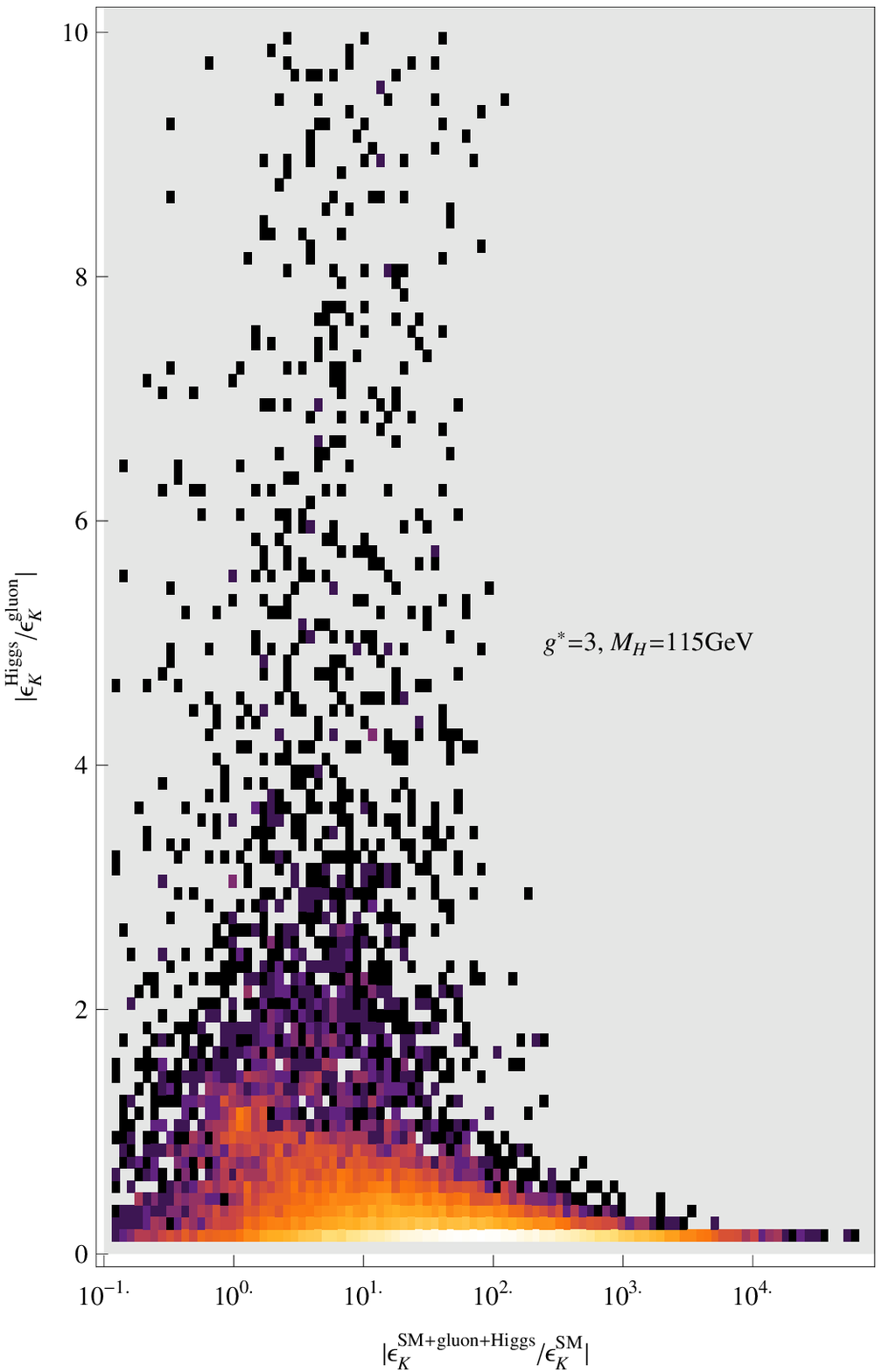}\includegraphics[height=.5\textwidth]{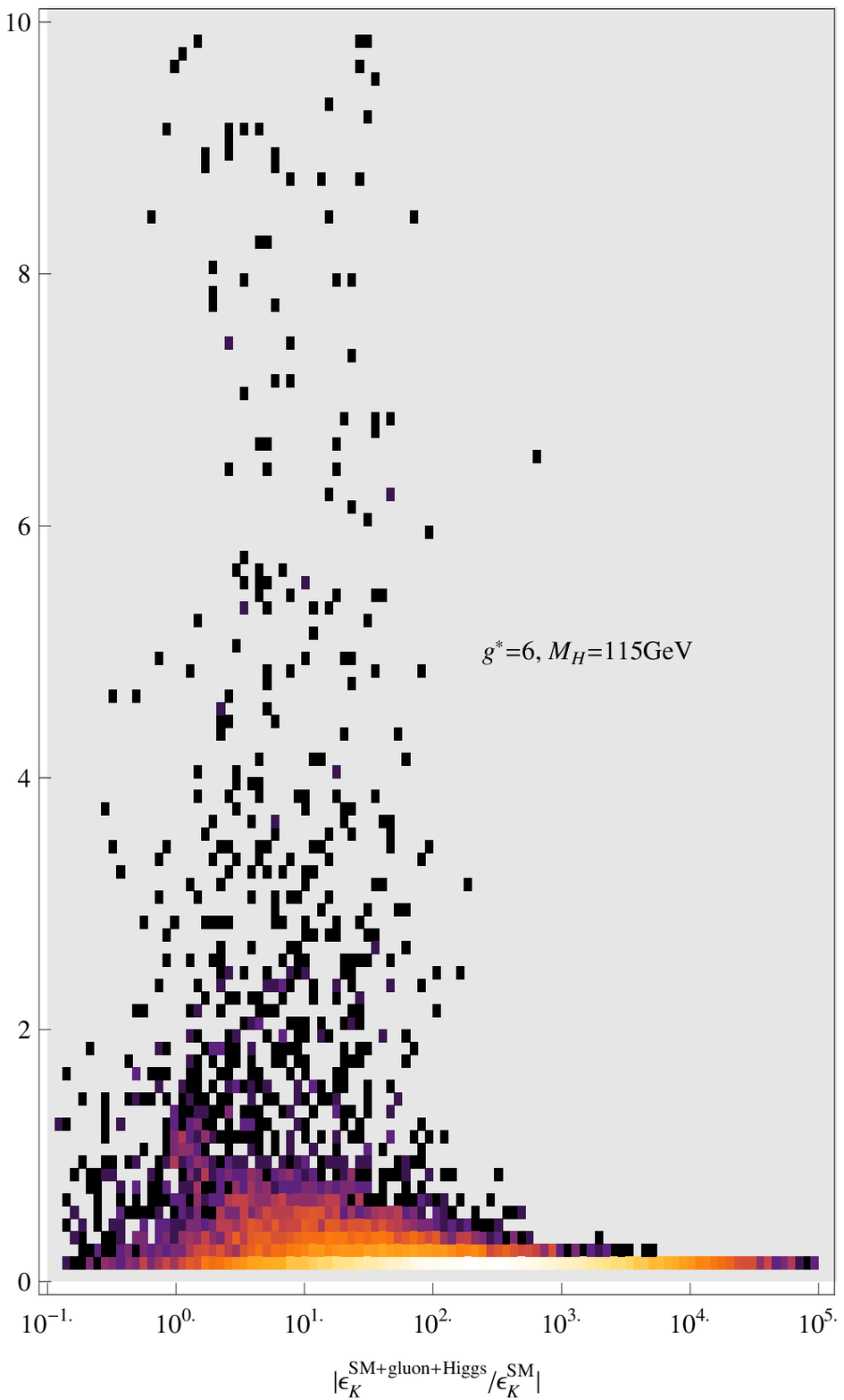}\includegraphics[height=.5\textwidth]{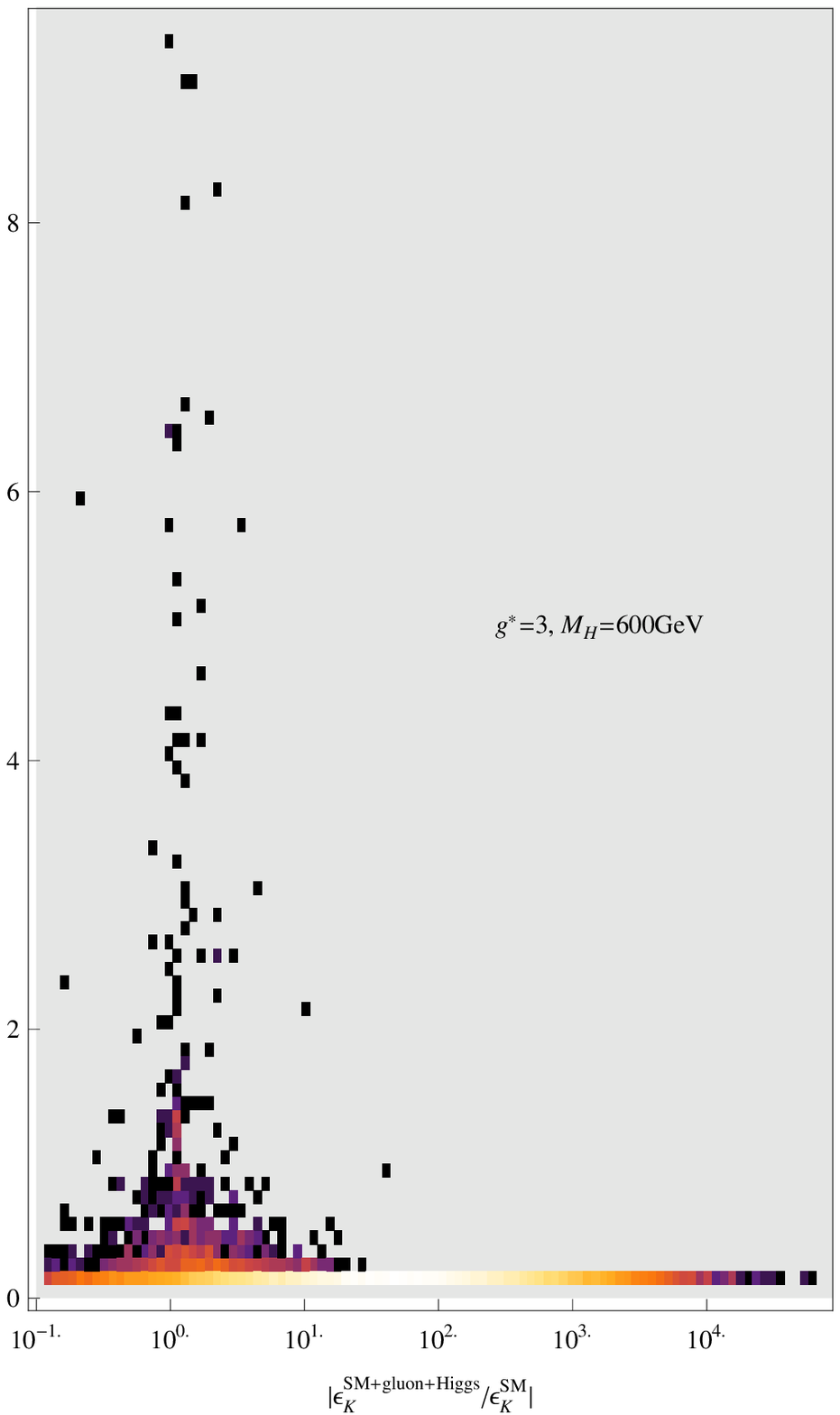}
\end{center}
\caption{Relative size of KK gluon and Higgs contributions to $\epsilon_K$ for $M_H=115\gev$ (left) and $M_H=600\gev$ (right). Light areas correspond to a high density of parameter points while dark areas correspond to a low density of parameter points in that region.
\label{fig:relative}}
\end{figure}

In Fig.~\ref{fig:relative} we show the relative sizes of the KK gluon
and Higgs contributions to $\epsilon_K$ for a KK scale $M_\text{KK}=2.45\tev$ for two values 
of the fundamental QCD coupling constant, $g^\ast=3$ and $6$, and for two values of the Higgs mass, $M_H=115\gev$ and $600\gev$.
In contrast to \cite{Blanke:2008zb} where the combined gauge boson correction to $\epsilon_K$ was considered we here give the purely gluonic contribution. However, as was shown in \cite{Blanke:2008zb}, the impact of electroweak (EW) gauge bosons to $K^0-\bar K^0$ oscillations is small compared to the impact of KK gluons.
{From the left panel of Fig.~\ref{fig:relative} we see that 
even in the case $(g^\ast,M_H)=(3,115\gev)$ in which the relative size of the Higgs contribution is maximal, most of the points in parameter space yield a much larger KK gluon contribution than Higgs contribution. For more than 65\% of all data points the Higgs contribution is smaller than 10\% of the KK gluon contribution. If the KK gluon contribution is accidentally small, it may well be exceeded by the Higgs contribution, however it is the data points that yield too large corrections to $\epsilon_K$ that will eventually set the bound on the KK mass scale. 

In Fig.~\ref{fig:ft} we show the average Barbieri-Giudice \cite{Barbieri:1987fn} fine-tuning of those points in parameter space that satisfy the $\epsilon_K$ constraint to $\pm30\%$ for the KK gluon case with $g^\ast=3$ (left) and for the Higgs case with $M_H=115\gev$ (right).
\begin{figure}[htbp]
\center{
\includegraphics[height=.3\textwidth]{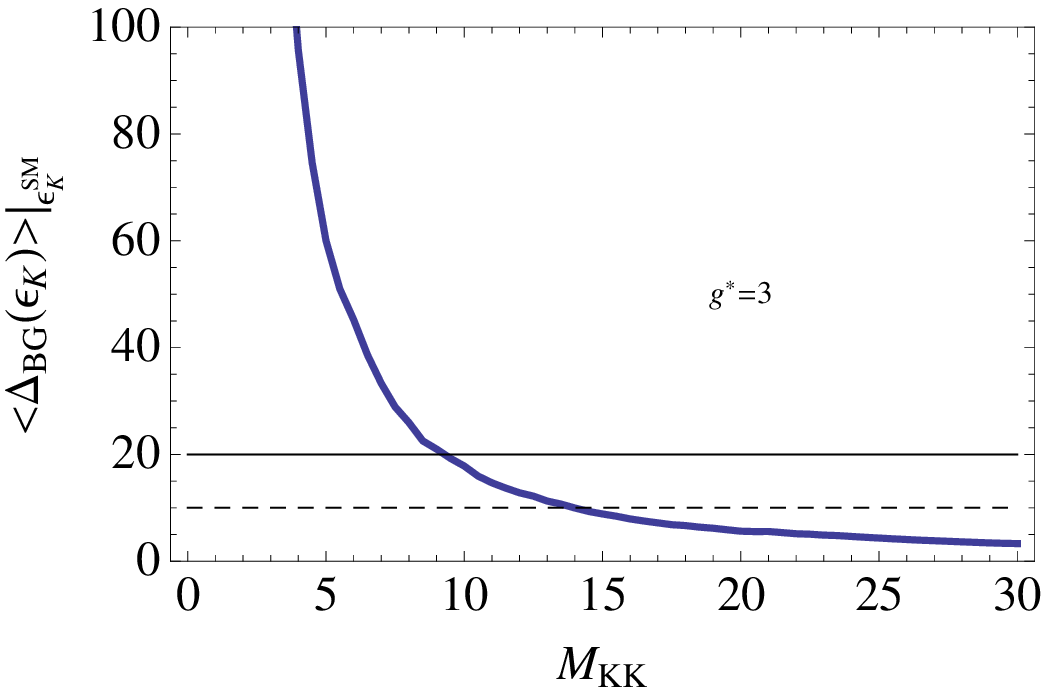}\hspace{1cm}\includegraphics[height=.3\textwidth]{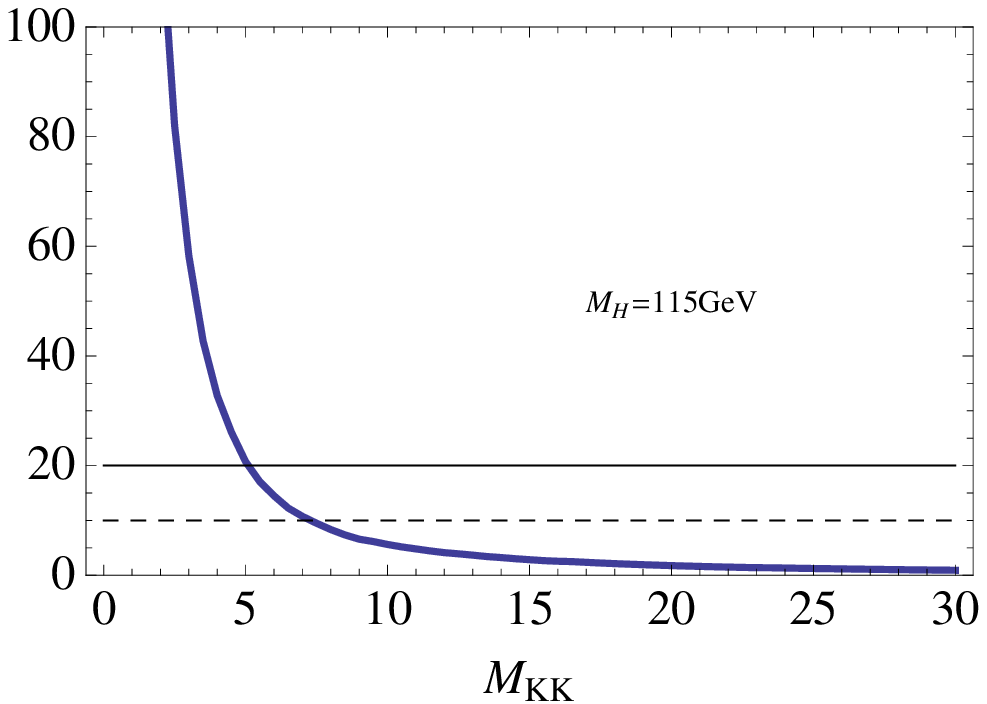}
}
\caption{The average required fine-tuning in $\epsilon_K$ as a function of the KK scale $M_\text{KK}$ for the gluonic contributions for $g^\ast=3$ (left) and for the Higgs contribution for $M_H=115\gev$ (right).\label{fig:ft}}
\end{figure}
From the left panel of Fig.~\ref{fig:ft} we can read off that depending on the amount of generic fine-tuning one is willing to accept a bound on the KK mass scale between $9\tev$ and $14\tev$ is required to keep the impact of KK gluon exchange under control (this is roughly half the value stated in \cite{Csaki:2008zd,Blanke:2008zb} for $g^\ast=6$). Under identical conditions the tree-level Higgs exchanges by themselves would imply a bound\footnote{Note that the bounds derived in this manner do not directly translate into the ratio $R_0$ which is introduced below.} on the KK mass scale between $5\tev$ and $7\tev$ for a Higgs mass $M_H=115\gev$, as can be seen from the right panel of Fig.~\ref{fig:ft}. For larger Higgs masses the bound gets accordingly weaker.

Taken together, Figs.~\ref{fig:relative} and \ref{fig:ft} show that in the brane Higgs scenario with $\hat Y_1^\text{max}=3$ (that roughly corresponds to $\langle\hat Y_1\rangle\simeq1.5$) the KK gluon contribution significantly exceeds the Higgs contribution for both $g^\ast=6$ and $g^\ast=3$. The tendency of this conclusion is not changed if the typical size of the Yukawa couplings $\langle\hat Y_1\rangle$ is raised to the maximal value that is compatible with the perturbativity estimate $\hat Y_1^\text{max}$.
So while in the brane Higgs case the tree-level exchange of the Higgs boson indisputably has an impact on $\epsilon_K$, the strongest bound on the KK mass scale for all values of the fundamental QCD coupling $g^\ast$ and all values of the Higgs mass comes from the exchange of KK gluons.\\

To extrapolate this finding to the bulk Higgs case we use \cite{Agashe:2008uz}
\begin{equation}
\left(\delta\epsilon_K\right)^\text{gluon}\propto\frac{\left(g^\ast\right)^2}{Y_\text{KK}^2}\frac{1}{a^2(\beta)}\frac{1}{M_\text{KK}^2}\,,\label{eq:gluon-contribution}
\end{equation}
where $a(\beta)$ depends on the localization of the Higgs field and $Y_\text{KK}$ is the KK fermion Yukawa coupling (in the brane Higgs scenario we would have $Y_\text{KK}=2\hat Y_1$). Furthermore, from
(\ref{eq:Delta-d-1}) we see that for the Higgs contribution to $\epsilon_K$
\begin{equation}
\left(\delta\epsilon_K\right)^\text{Higgs}\propto\frac{Y_\text{KK}^2}{M_\text{KK}^2}\,,\label{eq:Higgs-contribution}
\end{equation}
with nearly no dependence on the localization of the Higgs field \cite{Azatov:2009na}.

Using (\ref{eq:gluon-contribution}), (\ref{eq:Higgs-contribution}) we can infer an estimate for the ratio $R\equiv \left\langle(\delta\epsilon_K)^\text{gluon}/(\delta\epsilon_K)^\text{Higgs}\right\rangle$ for arbitrary values of $(g^\ast,Y_\text{KK},\beta)$ from the reference value $R_0$ that is determined for $(g^\ast=3,Y_\text{KK}\simeq3,\beta=\infty)$. Explicitly,
\begin{equation}
R(g^\ast,Y_\text{KK},\beta)\simeq\left(\frac{g^\ast}{3}\right)^2\left(\frac{0.5}{a(\beta)}\right)^2\left(\frac{3}{Y_\text{KK}}\right)^4R_0\,,
\end{equation}
where the ratio $R_0$ is found to be $R_0\sim 33$ and $a(\beta)$ is given for several values of $\beta$ in \cite{Agashe:2008uz}: $a(\infty)=0.5$, $a(2)=0.75$, $a(1)=1$, $a(0)=1.5$.

In Fig.~\ref{fig:R-ratio} we show the ratio \mbox{$R(g^\ast=3,Y_\text{KK},\beta)$} as a function of $\beta$ for two different values of $Y_\text{KK}$. The lower curve corresponds to the maximal value consistent with the perturbativity estimate, $Y_\text{KK}^\text{max}=6\sqrt{2}$ (where an additional factor $\sqrt{2}$ is due to the localization of the Higgs in the bulk \cite{Agashe:2008uz}), and the upper one to the value $Y_\text{KK}=1/2Y_\text{KK}^\text{max}=6/\sqrt{2}$, that corresponds to the average if values are randomly chosen between $0$ and the maximal value.
We observe that as soon as the Higgs field is detached from the IR brane the Higgs contribution to $\epsilon_K$ can in principle exceed the KK gluon contribution, although depending on the typical size of Yukawa couplings this outcome is not imperative. The possible dominance of the Higgs contributions is largely due to the increase of the maximally allowed value for $Y_\text{KK}$ by a factor of $\sqrt{2}$, but also by the shift of the quark zero modes towards the UV brane that becomes possible for a bulk Higgs.
\begin{figure}[htbp]
\center{
\includegraphics[width=.5\textwidth]{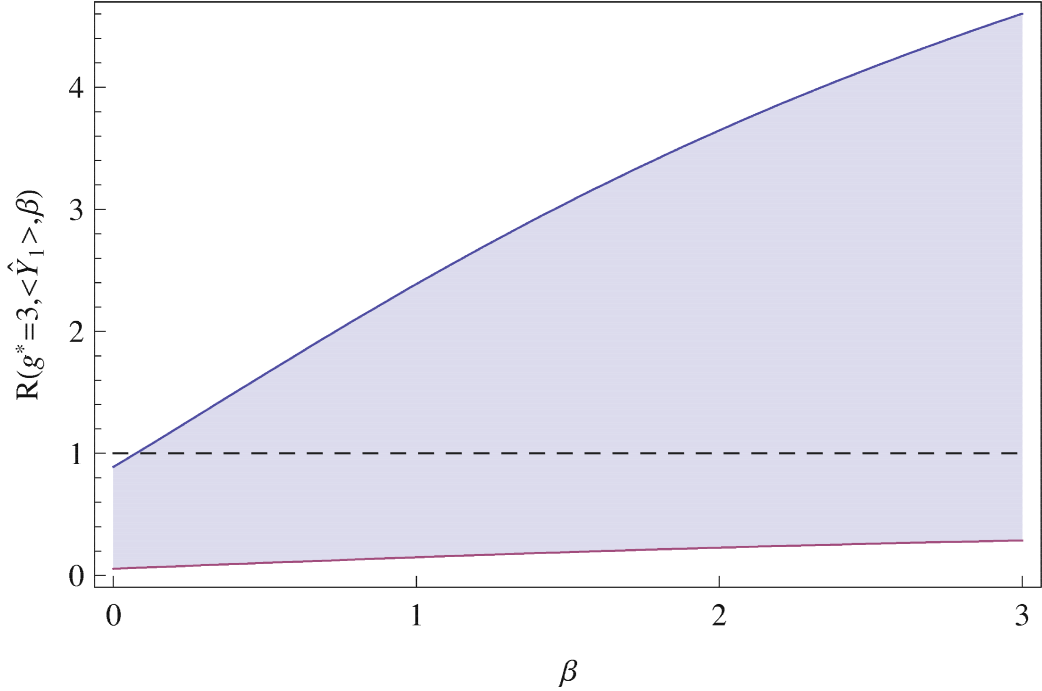}
}
\caption{The ratio R for $g^\ast=3$ as a function of $\beta$ for $Y_\text{KK}=6\sqrt{2}$ (lower curve) and $Y_\text{KK}=6/\sqrt{2}$ (upper curve).\label{fig:R-ratio}}
\end{figure}

At this point it is important to keep in mind that observables that depend on positive powers of $Y_\text{KK}$, such as the neutron EDM $d_n$ \cite{Agashe:2004cp}, $Br(B\to X_s\gamma)$ \cite{Agashe:2008uz} and $\epsilon^\prime/\epsilon$ \cite{Gedalia:2009ws} for fixed $M_\text{KK}$ constrain the size of the Yukawa couplings such that the above statements are only sensible if the isolated $\epsilon_K$ constraint is considered.

\newsection{Conclusions}\label{sec:Conclusions}
In this paper we have carefully compared the impact of KK gluon and Higgs exchanges on the observable $\epsilon_K$ in the brane Higgs scenario.
Subsequently we have extrapolated our results to the case of a bulk Higgs and given estimates for the relative size of KK gluon and Higgs contributions to $\epsilon_K$.
A comparison of KK gluon and Higgs contributions to $\epsilon_K$ is mandatory since up to now those two contributions were never treated simultaneously and the generic bounds on the KK scale deduced from their presence hence depend on different and often implicit assumptions. These assumptions for instance include the uncertainties within which an observable is required to be reproduced or the individual acceptance of fine-tuning.

In the course of our analysis we have shown in detail how the value of the low energy observable $\epsilon_K$ can be derived from the misalignment of quark mass matrices and Yukawa couplings at the KK scale for which analytic expressions were derived in \cite{Azatov:2009na}. Of particular importance in this context is the proper employment of the RG equations. While the Wilson coefficients of the effective interactions induced by KK gluon exchange have to be run down from the KK scale to the physical scale according to their anomalous dimension, this is different for the case of the Higgs induced contributions. Here the anomalous Higgs couplings that are induced at the KK scale have to be evolved down to the scale of the Higgs boson mass where new interactions are generated by the exchange of the Higgs boson. From there, as in the KK gluon case, the Wilson coefficients of the effective interactions have to be evolved to the physical scale according to their anomalous dimensions.

As an outcome of our analysis we have shown that while Higgs FCNCs have an impact on $\epsilon_K$ as was already pointed out in \cite{Azatov:2009na}, their contribution for a brane Higgs is dwarfed by the contribution of KK gluons even in the most favorable scenario for the Higgs contributions. For a brane Higgs scenario with reasonable choices fundamental Yukawa couplings the bound on the KK scale implied by the presence of tree-level Higgs exchanges is found to be $5-7\tev$, to be compared to the 
corresponding bounds for KK gluon exchange that are given by $M_\text{KK}\gtrsim(9\tev-14\tev)$ for $g^\ast=3$ and $M_\text{KK}\gtrsim(19\tev-32\tev)$ for $g^\ast=6$ (see also \cite{Csaki:2008zd,Blanke:2008zb,Duling:2009sf}).

Tentatively extrapolating this result to the case of a bulk Higgs we find that the Higgs contribution to $\epsilon_K$ can exceed the KK gluon contribution as soon as the Higgs field is detached from the IR brane if the Yukawa couplings are assumed to have the maximal value still consistent with perturbativity estimates. This is largely due to larger allowed values for the Yukawa couplings in the presence of a bulk Higgs. For Yukawa couplings smaller by a factor of two than the maximal value the KK gluon contributions are typically larger than the Higgs contributions for Higgs localizations $\beta$ down to $\beta\simeq0$.

Finally we would like to mention that in the present work we only studied the isolated constraint on the KK scale arising from the observable $\epsilon_K$. The total bound on $M_\text{KK}$ is generally higher, as also other observables such as $d_n$ \cite{Agashe:2004cp}, $Br(B\to X_s\gamma)$ \cite{Agashe:2008uz} and $\epsilon^\prime/\epsilon$ \cite{Gedalia:2009ws} impose constraints on the KK scale. The RS contributions to these observables are proportional to $Y_\text{KK}^2$ such that for values of $Y_\text{KK}$ as large as the perturbativity bound the strongest  individual bound on $M_\text{KK}$ typically comes from on of those observables instead of $\epsilon_K$.

\subsubsection*{Acknowledgments}
\noindent 
I thank Andrzej Buras for very helpful discussions and his critical reading of the manuscript. I also thank Kaustubh Agashe for his comments and enlightening discussion as well as Alexandr Azatov, Manuel Toharia, Andreas Weiler and Lijun Zhu for their comments. This research was partially supported by the Graduiertenkolleg GRK 1054 of Deutsche Forschungsgemeinschaft (DFG).

\begin{appendix}
\end{appendix}

\addcontentsline{toc}{chapter}{References}
\providecommand{\href}[2]{#2}\begingroup\raggedright\endgroup

\end{document}